\documentclass[11pt,twoside]{article} 
\usepackage{cspm-asp2006}
\usepackage{epsfig,graphicx,natbib,url}
\usepackage{lscape} 
\begin{document}
 \setcounter{page}{573}

\markboth{Hammerschlag et al.}{Options for the DOT}
\title{Aperture Increase Options for the Dutch Open Telescope}
\author{R.\,H.\,Hammerschlag, F.\,C.\,M.\,Bettonvil, A.\,P.\,L.\,J\"agers and R.\,J.\,Rutten}

\affil{Sterrekundig Instituut, Utrecht University, The Netherlands}

\begin{abstract}
  This paper is an invitation to the international community to participate
  in the usage and a substantial upgrade of the Dutch Open Telescope
  on La Palma (DOT, \url{http://dot.astro.uu.nl}).

  We first give a brief overview of the approach, design, and current
  science capabilities of the DOT.  It became a successful
  0.2-arcsec-resolution solar movie producer through its combination
  of {\em (i)\/} an excellent site, {\em (ii)\/} effective wind flushing
  through the fully open design and construction of both the 45-cm
  telescope and the 15-m support tower, {\em (iii)\/} special designs
  which produce extraordinary pointing stability of the tower,
  equatorial mount, and telescope, {\em (iv)\/} simple and excellent
  optics with minimum wavefront distortion, and {\em (v)\/}
  large-volume speckle reconstruction including narrow-band
  processing.  The DOT's multi-camera multi-wavelength speckle imaging
  system samples the solar photosphere and chromosphere simultaneously
  in various optical continua, the G band, Ca\,II\,H (tunable
  throughout the blue wing), and H$\alpha$ (tunable throughout the
  line).  The resulting DOT data sets are all public.  The DOT
  database (\url{http://dotdb.phys.uu.nl/DOT}) now contains many
  tomographic image sequences with 0.2-0.3~arcsec resolution and up to
  multi-hour duration.  You are welcome to pull them over for
  analysis.

  The main part of this contribution outlines DOT upgrade designs
  implementing larger aperture.  The motivation for aperture increase
  is the recognition that optical solar physics needs the
  substantially larger telescope apertures that became useful with the
  advent of adaptive optics and viable through the DOT's open
  principle, both for photospheric polarimetry at high resolution and
  high sensitivity and for chromospheric fine-structure diagnosis at
  high cadence and full spectral sampling.

  Our upgrade designs for the DOT are presented in an incremental
  sequence of five options of which the simplest (Option~I) achieves
  1.4~m aperture using the present tower, mount, fold-away canopy, and
  multi-wavelength speckle imaging and processing systems.  The most
  advanced (Option~V) offers unblocked 2.5~m aperture in an off-axis
  design with a large canopy, a wide 30-m high support tower, and
  image transfer to a groundbased optics lab for advanced
  instrumentation.  All five designs employ adaptive optics.  The
  important advantages of fully open, wind-transparent and
  wind-flushed structure, polarimetric constancy, and absence of
  primary-image rotation remain.  All designs are relatively cheap
  through re-using as much of the existing DOT hardware as possible.

  Realization of an upgrade requires external partnership(s). This
  report about DOT upgrade options therefore serves also as initial
  documentation for potential partners.
\end{abstract}

\section{Introduction}

\begin{figure}
  \centering \includegraphics[width=\textwidth]{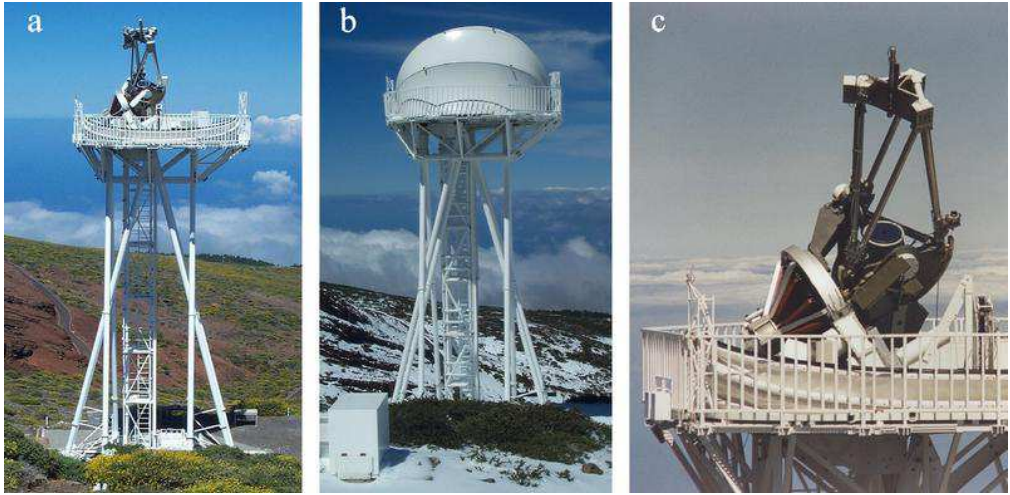}
  \caption[]{\label{rr-rh-fig1:existing DOT}
  The existing DOT with 45-cm primary mirror at 2350~m altitude on La
  Palma.  \\{\em (a)\/} The DOT in operation. The 15-m tower and the
  telescope are sufficiently transparent to not disturb the wind which
  maintains temperature homogeneity in and around the telescope.  At
  sufficient wind strength (7~km/h can be enough depending on the wind
  direction) the larger temperature fluctuations occurring near ground
  level do not reach the telescope. The special tower geometry keeps
  the platform parallel to the ground even under strong wind loads.
  The clam-shell canopy is opened completely for observations.
  \\{\em (b)\/} When not in operation the telescope is protected by
  closing the folding canopy.  It is made of strong tensioned
  polyester cloth with an outer PVDF coating on which snow and ice do
  not stick. The canopy can be opened and closed within a few minutes
  in winds up to 100~km/h. When closed it can withstand much stronger
  winds, and has already survived storms of 200~km/h.  \\{\em (c)\/}
  The telescope close-up. The primary mirror and the optical beam to
  the primary focus are fully open to wind.  The DOT was the first
  telescope showing that such an open air path can permit
  diffraction-limited resolution.  Note that the primary mirror is
  located well above the declination axis of the equatorial mount.  It
  sticks out high above the platform into full wind flushing.
}
\end{figure}

During the late 1990s the Dutch Open Telescope (DOT,
Fig.~\ref{rr-rh-fig1:existing DOT} and \url{http://dot.astro.uu.nl})
was the pioneering demonstrator of the open-telescope technology now
pursued in the German GREGOR and BBSO NST projects and inspirational
to the US ATST project.  These projects capitalize on the advents in
wavefront restoration through adaptive optics (AO) and numerical image
processing which now enable meter-class image sharpness, far beyond
the best Fried-parameter values at any site, and so require telescope
technology beyond the 1-m evacuated-telescope technology limit
realized by the Swedish 1-m Solar Telescope (SST).  In the meantime
the 45-cm DOT became an outstanding supplier of solar-atmosphere
movies sampling the photosphere and chromosphere simultaneously at up
to 0.2~arcsec resolution (Section~\ref{dot-sec:summary}).  The
resulting image sequences are publicly available for analysis
(Section~\ref{dot-sec:database}).

The major science drivers for aperture increase beyond the SST are:

\begin{enumerate} \vspace*{-1ex} \itemsep=0ex

  \item {\em Photosphere\/}: precise, deep, and complete Stokes
        polarimetry at high angular resolution, preferably combining
        visible and infrared lines in 2-D mapping.
        Targets: umbrae, penumbrae, pores, plage
        and network magnetic elements, internetwork fields, etc.

  \item {\em Chromosphere\/}: high to very high cadence
        profile-sampled narrow-band imaging in chromospheric lines, in
        particular Ca\,II\,H\&K with spectral sampling throughout the
        extended line wings in order to follow dynamic phenomena with
        height throughout the upper photosphere and low chromosphere,
        and the Ca\,II infrared lines and the Balmer lines with full
        profile mapping in order to disentangle the complex opacity,
        source function, and Doppler sensitivities that make the
        chromosphere such a rich scene in these lines.  Targets:
        filaments, active regions, mottles/fibrils/spicules, Ellerman
        bombs, flares, wave patterns, shock dynamics, etc.

\end{enumerate} \vspace*{-1ex}
For both, sufficient photon collection is the principal large-aperture
motivation.  To freeze the seeing no exposure should exceed 10\,ms,
also for narrow-band diagnostics.  Multi-frame image collection for
MOMFBD restoration
  (\cite{rh-2005SoPh..228..191V}) 
or many-frame collection for speckle reconstruction must be completed
within the solar-change time per resolution element, including
spectral profile sampling as necessary.  The change time becomes
shorter for larger angular resolution and can be much shorter than the
sound-speed crossing time, as demonstrated by the recent 1-s-cadence
H\hspace{0.1ex}$\alpha$\ movies of
         \citet{rh-2006ApJ...648L..67V}.

These science motivations plus the fact that the 50-cm SOT onboard
Hinode duplicates many current DOT capabilities at a much higher duty
cycle (no bad seeing, no bad weather, no nights during its
non-eclipse seasons) led us to the larger-aperture designs discussed
below.  In Section~\ref{dot-sec:upgrades} we outline how conversion of
the 45-cm DOT into a larger solar telescope is not only feasible but
actually a relatively cheap venue to meter-class angular
resolution through using existing parts where possible.  We focus on
1.4-m and 2.5-m strawman designs in a sequence of options ranging from
minimum cost to maximum science capability.

We cannot realize such DOT upgrades on our own; they require external
support.  At present the DOT is run on a budget of about
250~kEuro/year covering salaries (excluding the first and last author
who are academic staff), travel to La Palma and equipment, plus an
additional allocation of up to 3000~manhours/year of Utrecht
University workshop effort encompassing mechanical design and
fabrication as well as electronics and software development in very
close collaboration with the DOT team.  Our present funding covers
these costs up to 2008, but not beyond that date while a larger budget
is required to realize and operate a larger telescope
(Section~\ref{sec:costs}).  This paper is therefore intended as
initial documentation for potential partners.

\section{DOT Technical Overview}   \label{dot-sec:summary}

The DOT performs so well thanks to the combination of {\em (i)\/}
its wind-swept oceanic mountain site at the Observatorio del Roque
de los Muchachos on the Canary Island La Palma, {\em (ii)\/} minimum
obstruction to the wind by the very open tower, the very open
telescope, and the fully-folding canopy, {\em (iii)\/} effective
wind-flushing of the open telescope, {\em (iv)\/} short-exposure
speckle imaging, and {\em (v)\/} the consistent application of
speckle restoration in an on-site processor farm.

The DOT design and construction are characterized by rigorous
adherence to its open principle and large emphasis on mechanical
stability.  The open tower, fold-away canopy, and equatorial telescope
mount are highly transparent to the fairly laminar Northern trade
winds that bring the best seeing at the Roque de los Muchachos. They
don't spoil the seeing; in addition, the wind flushes the telescope
interior faster than internal turbulence can develop.  The remaining
higher-layer wavefront aberrations are corrected through speckle
processing.  Its advantages are that it restores the full field of
observation in equal measure and that it delivers rather good results
already at relatively poor seeing.  It requires a large amount of
post-processing but this has been remedied with the parallel DOT
Speckle Processor in a nearby building.  The complete system delivers
0.2~arcsec diffraction-limited image quality whenever the seeing is
reasonable, already at Fried-parameter values of order 6-10\,cm.  At La
Palma such seeing sometimes occurs during multiple hours.

During the past years the DOT has been equipped with an elaborate
multi-wavelength imaging system harboring six identical speckle
cameras that register wide-band continua in the blue and red, the G
band at 4305\,\AA, Ca\,II\,H\ with an interference filter that can be
tuned per speckle burst through the blue line wing, and narrow-band
H\hspace{0.1ex}$\alpha$\ using a 250~m\AA\ FWHM Lyot filter that can also be tuned per
speckle burst.  Two-channel speckle reconstruction following
  \citet{rh-1992A&A...261..321K} 
permits the registration of multi-wavelength H\hspace{0.1ex}$\alpha$\ movies at
20-30~s cadence or single-wavelength H\hspace{0.1ex}$\alpha$\ movies at much faster
cadence.

The DOT is usually manned from early spring until late autumn, usually
with the first author taking care of the telescope operation and
P.~S\"utterlin in control of all observing and speckle processing.  A
typical two-week campaign delivers on average 5--6 days with good
data.  Thanks to the parallel processing the data now become available
soon after the campaign.

More detail is given in
  \citet{rh-DOT-Hammerschlag1981a}
for the original DOT design description,
  Rutten et al.\ (2004a, 2004b)
  \nocite{rh-DOT-StPetersburg2004}
  \nocite{rh-DOT-tomo1} 
for general overviews,
  \citet{rh-DOT-2006SPIE.6273E..50H} 
for the tower design,
  \citet{rh-DOT-2006SPIE.6269E..12B} 
for the multi-wavelength imaging,
  \citet{rh-DOT-SPW4} 
for ongoing work on Ba\,II\,4554 polarimetry,
and
  \citet{rh-DOT-dot++2004} 
for an earlier description of a DOT upgrade to 1.4~m aperture. All
DOT papers are available at
\url{http://dot.astro.uu.nl}.

\section{DOT Database} \label{dot-sec:database}

All DOT data are public.  The DOT has collected high-resolution movies
of the sun since the autumn of 1999 in an increasing number of
spectral diagnostics.  The yearly harvest increased markedly in 2005
with the advent of the large-volume parallel DOT Speckle Processor.
The DOT database resides at ftp site \url{ftp://dotdb.phys.uu.nl/} and
has a user-friendly graphical interface at
\url{http://dotdb.phys.uu.nl/DOT/} which for every day with worthwhile
data serves a thumbnail pictorial index of what was collected.  It
also specifies the target, observing mode, time of observation,
cadence, solar disk location, average seeing quality (Fried parameter
$r_0$), a link to the pertinent Mees active region map, a
burst-by-burst plot of the Fried parameter, and a ``Get data'' link to
the corresponding directory in the DOT database.  The basic DOT data
product consists of FITS files per speckle-reconstructed image, but
for many runs the data base also serves processed and aligned image
sequences as data cubes.

You are welcome to use any DOT data for detailed analysis.  Explanation
and IDL reading instruction is given under ``DOT data'' at
\url{http://dot.astro.uu.nl}.  Questions should go to
P.Suetterlin@astro.uu.nl.

We plan to integrate the DOT data base into the Virtual Solar
Observatory (\url{http://sdac.virtualsolar.org}) and to add more
search capability as well as other user commodities.

\begin{figure}
  \centering
  \includegraphics[width=\textwidth]{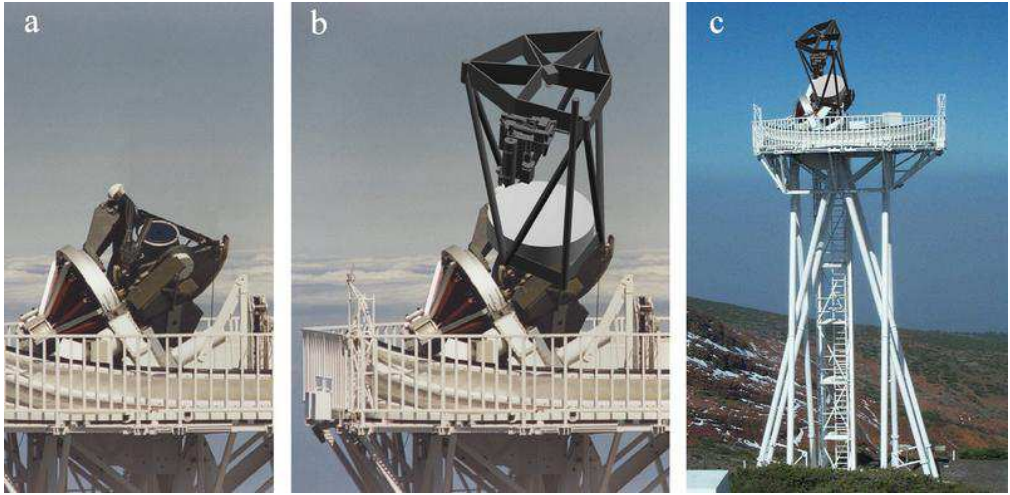}
  \caption[]{\label{rr-rh-fig2:upgrade1.4}
  Option I: DOT upgrade to 1.4-m aperture.  \\{\em (a)\/} Remove the
  existing telescope top.  \\{\em (b)\/} Place a new mirror with a new
  mirror support on the existing equatorial mount, add a new telescope
  top, relocate the existing multi-channel imaging system to its side,
  and add a new prime-focus optics package.  The optical scheme is
  shown in Fig.~\ref{rr-rh-fig3:optics1.4}.  \\{\em (c)\/} The result
  -- in observing position -- on top of the existing tower.  At a
  focal length of $f=2.3$\,m the telescope can still be parked within
  the existing 7-m canopy.
}
\end{figure}

\begin{figure}
  \centering
  \includegraphics[width=11cm]{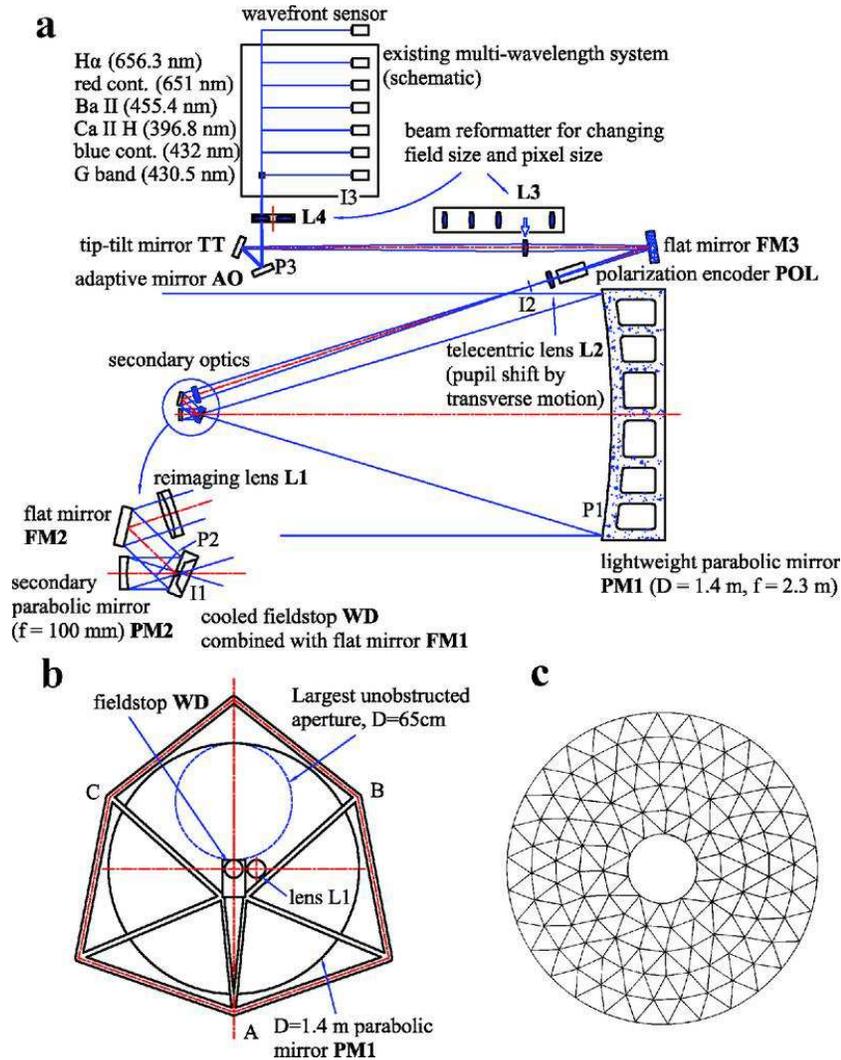}  
  \caption[]{\label{rr-rh-fig3:optics1.4}
  Optical layout of Option~I with 1.4-m primary mirror.  \\{\em (a)\/}
  Side view.  The parabolic primary and secondary mirror together are
  coma-free to produce diffraction-limited quality over the full field
  of view.  L1 produces an enlarged image I2 near the field lens
  L2. The ensembles of interchangeable lenses L3 and L4 produce
  user-selectable choice between angular resolution and field size
  whereas transverse translation of telecentric field lens L2 offers
  transversal pupil shift. \\{\em (b)\/} Top view of the aperture =
  primary mirror with spider shadow.  Transversal pupil shift allows
  obstruction-free apertures up to 65\,cm (circle in the upper part).
  The inclined flat mirrors FM1 and FM2 reflect the light in
  perpendicular directions to compensate partial polarization.  {\em
  (c)\/} Alternative spider design using very thin plates gradually
  spreading in all directions to obtain a diffraction pattern without
  pronounced ghosts.  Such spider geometry can be optimized together
  with the AO layout.
}
\end{figure}

\section{DOT Upgrade Designs}  \label{dot-sec:upgrades}
In the remainder of this contribution we present a range of options
to increase the DOT aperture to meterclass size, in order of increasing
gain in science capability and cost.

\subsection{Option~I: upgrade to 1.4-m aperture}  \label{dot-sec:upgrade1.4m}
The simplest, Option~I, is to upgrade the existing DOT with 45-cm
primary mirror shown in Fig.~\ref{rr-rh-fig1:existing DOT} to 1.4-m
aperture by removing the existing telescope top, placing a new
mirror support with a new 1.4-m mirror with focal length $f=2.3$\,m
(opening ratio $f/1.64$) on the existing telescope mount, adding a
new prime-focus optics package (secondary parabolic mirror, two flat
mirrors and re-imaging lens) and adaptive optics, and constructing a
new telescope-top support structure
(Figs.~\ref{rr-rh-fig2:upgrade1.4} and \ref{rr-rh-fig3:optics1.4}).
Re-used are the existing tower, the platform, the folding canopy,
the equatorial mount, the multi-wavelength imaging system, and the
image acquisition and processing computers in the nearby SST and
Automatic Transit Circle (ATC) buildings. We call this simplest and
cheapest upgrade Option~I here. It was called DOT$^{++}$ in a
proposal detailed in
\citet{rh-DOT-dot++2004}
and is summarized here.

The optical layout is shown in Fig.~\ref{rr-rh-fig3:optics1.4}.  Both
the primary and secondary mirrors are parabolic because two parabolic
mirrors in cascade leave no coma, making the whole field diffraction
limited.  The two 20\deg\ inclined flat mirrors FM1 and FM2 reflect
in mutually perpendicular directions compensating their partial
polarization.  The layout includes a tip-tilt mirror and an
adaptive-optics mirror.  The spider construction may be optimized
commensurate with the AO pupil geometry
(Fig.~\ref{rr-rh-fig3:optics1.4}c).

This optical design permits a flexible near-instantaneous user choice
between angular resolution and field size, achieved with the sets of
interchangeable ``zoom-out'' lenses L3 and L4 and maintaining the photon
flux per pixel.  Transversal pupil shift is possible through
translation of telecentric field lens L2 and enables selection of
obstruction-free apertures up to 65\,cm. See
 \citet{rh-DOT-dot++2004}
for detail.

An advantage of the equatorial mount is that there are no image
rotations relative to the AO-system for the optics system on the
pointed telescope structure.  Combination of AO with post-detection
numerical wavefront restoration is possible and desirable.

\subsection{Option~II: upgrade to 2.5-m aperture}  \label{dot-sec:upgrade2.5m}

\begin{figure}
  \centering
  \includegraphics[width=\textwidth]{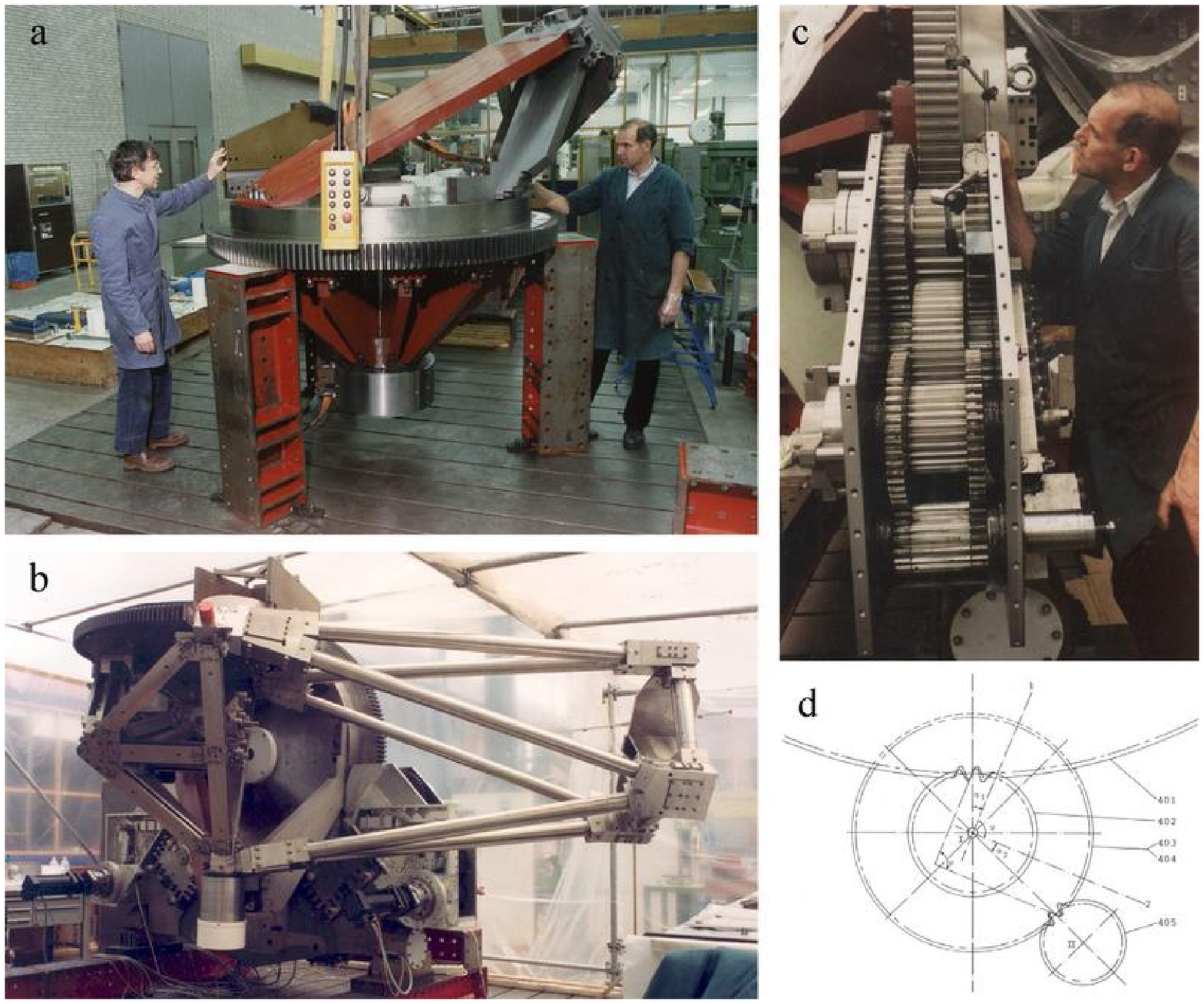}
  \caption[]{\label{rr-rh-fig4:mount+gears}
  Fabrication of the exceedingly stable DOT mount and drives.  \\{\em
  (a)\/} Hour-angle gear wheel (190\,cm diameter) with above the teeth
  the roller raceway for the radial support. The fork is in
  construction above the raceway. Below, near the floor, is the large
  double-row spherical roller bearing which was custom-made by SKF to
  minimize clearance. It has a large central hole to permit beam
  passage to an optional optics lab.  \\{\em (b)\/} The mount and
  drives fully assembled in the workshop, with the hour-angle axis
  oriented horizontally.  The mount permits pointing in all
  directions.  {\em (c)\/} One of the two hour-angle gears.  All gears
  have self-aligning pinions. This special design, a DOT invention to
  minimize the risk of stick-slip, achieves line contact between the
  meshing teeth over the whole tooth width even under the relatively
  small loads used here.  \\{\em (d)\/} Principle of the self-aligning
  pinions. On both sides of a pinion is a gear of the next stage. This
  pair of gears is fixed with a ring of screws to the pinion. It forms
  a single block, which is supported by a single double-row spherical
  roller bearing. This bearing permits rotation around three axes: the
  normal rotation around axis $I$ perpendicular to the drawing plane,
  and two perpendicular axes $1$ and $2$ in the plane of the drawing.
  Rotations of the block around axis $1$ provide line contact of the
  pinion with the large gear wheel whereas rotation around axis $2$
  provides line contact between the gears of the next stage (pinion
  II). The two half parts of the next gear were ground together
  for optimum matching.
}
\end{figure}

The existing canopy has a diameter of 7~m which limits the focal
length of the primary mirror to 2.3~m.  This limits the aperture
diameter to 1.4~m because a faster opening ratio brings too severe
optical and thermal problems.

However, the existing equatorial DOT telescope mount and drives are
sufficiently stable, stiff, and large to harbor a 2.5~m diameter
mirror with appropriate mirror support and prime-focus support
structures without modification.  Fig.~\ref{rr-rh-fig4:mount+gears}
shows photographs of the extraordinary stable DOT mount and gears.
They were considerably overdimensioned in the original design to
ensure strict rigidity and enable the option of installing a larger
mirror later.  The equatorial mount has a declination gear wheel with
pitch circle diameter 174\,cm, pitch 2.5\,cm and teeth width 7\,cm and
an hour-angle gear wheel with pitch circle 190\,cm, pitch 3.1\,cm and
teeth width 11\,cm.  The DOT drives are not only overdimensioned but
also employ self-aligning pinions invented, developed, and tested
during the DOT construction.  Their principle is illustrated in
Fig.~\ref{rr-rh-fig4:mount+gears}d; more details are given in
\citet{rh-DOT-Hammerschlag1983}.
These pinions achieve line contact between meshing teeth over the
whole tooth width under typical telescope loads.  The latter are small
compared with standard mechanical practice.  Under such small loads
teeth contact in classical drives occurs only over a part of the tooth
width, reducing the structural stiffness.  Large preloads are no
remedy since they would produce stick-slip irregularities in the very
slow motion needed to follow targets. These self-aligning pinions
turned out to work exceptionally well, so that the gears are actually
much stiffer than the design tolerance of deformations $\leq
0.07$\,arcsec at wind speeds 0--10~m/s set by the DOT diffraction
limit.

\begin{figure}
  \centering
  \includegraphics[width=\textwidth]{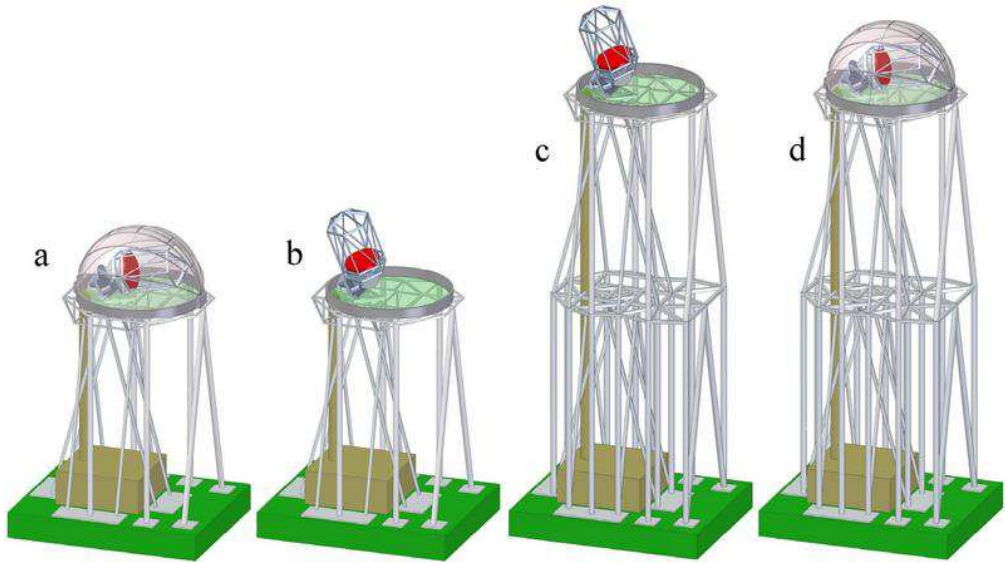}
  \caption[]{\label{rr-rh-fig5:upgrade2.5}
  Options II and III with 2.5-m primary mirror.
  \\{\em (a)\/} Option II: enlarged 15-m tower and platform,
  carrying 9~m canopy with 2.5-m mirror on the existing mount. \\ {\em
  (b)\/} Option II with open canopy and telescope in operating
  position.
  \\{\em (c)\/} Option III: the Option~II tower with platform, telescope
  and canopy put on top of an secondary base structure of 15-m height,
  bringing the telescope to 30~m above the ground.  The added base
  structure also has a geometry which maintains parallel motion of the
  platform relative to the ground under varying wind loads.  Such
  parallel motion combines well with the geometry of the upper tower
  with separated base points of the triangles (see text for further
  explanation). \\{\em (d)\/} Option III with telescope in parked
  position and closed canopy.
}
\end{figure}

\begin{figure}
  \centering
  \includegraphics[width=\textwidth]{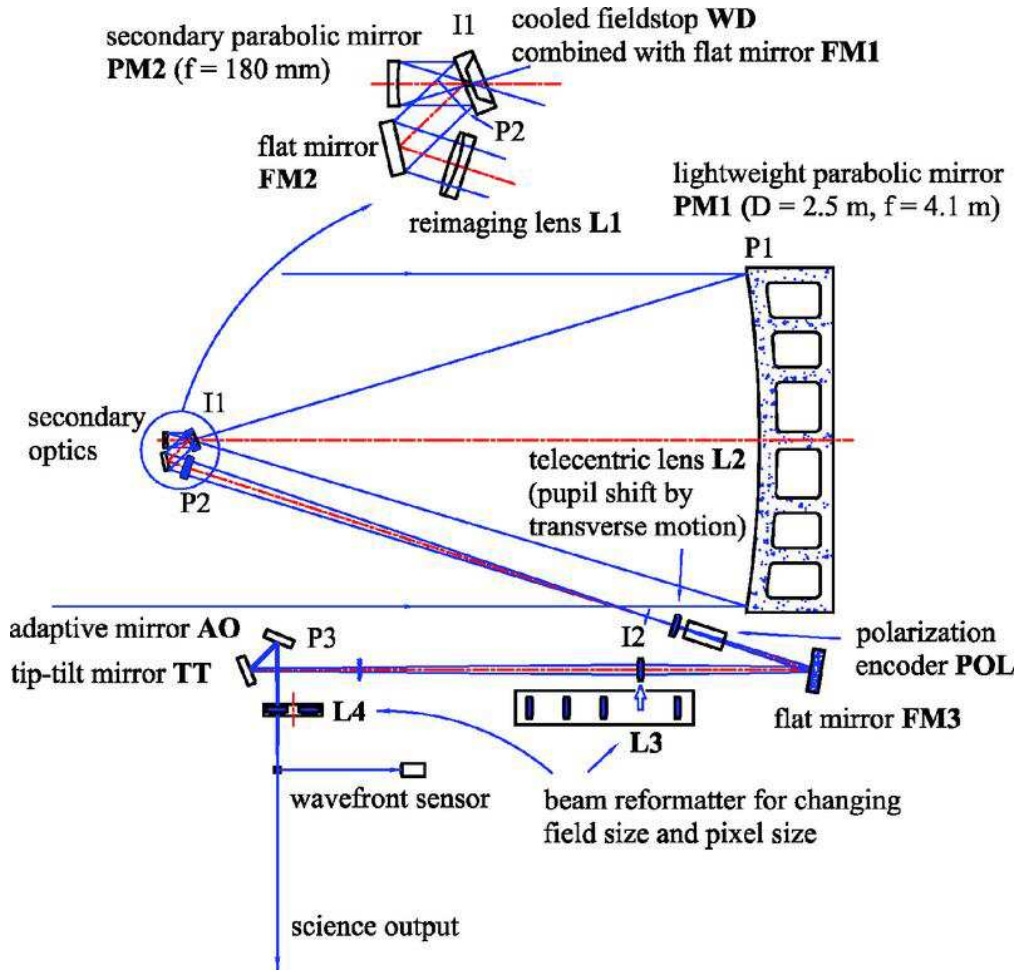}
  \caption[]{\label{rr-rh-fig6:optical layout II-IV}
  Optical
  layout for the options II to IV with 2.5-m mirror, similar to the
  1.4-m upgrade with the same advantages.
}
\end{figure}

Thus, for an upgrade to a 2.5-m mirror the DOT mount can simply stay
but a larger canopy is needed for bad-weather telescope protection.
A telescope with 4.1~m focal length will just fit within a canopy of
9~m diameter as illustrated in Fig.~\ref{rr-rh-fig5:upgrade2.5}a.
This is the size of the canopy for the GREGOR telescope on Tenerife
which was installed by us in 2004 and already survived an
exceedingly strong storm with gusts of 245~km/hr.
More detail is given in
  \citet{rh-DOT-Hammerschlag++Beijing2006}.

Such size increase also requires an enlargement of the platform
including additional support from the ground to make the enlarged
platform stiff enough against wind loading.  The latter support is
shown in Fig.~\ref{rr-rh-fig5:upgrade2.5}a,b.  It consists of an
isosceles triangle on the north side and four vertical posts -- two on
the east side and two on the west side.  Like the already existing
4~isosceles triangles of the present DOT tower, the additional
elements provide a geometry that keeps the platform parallel to the
ground at the small but inevitable leg deformation under wind
load. More information about geometries for such parallel motion is
given by
  \citet{rh-DOT-2006SPIE.6273E..50H}.
The additional triangle also increases the stiffness against platform
rotations around a vertical axis.

The diameter of the additional tubes is set at 406.4~mm, whereas the
diameter of the tubes making up the existing tower is 244.5~mm. The
choice of a larger diameter over the same free distance of 15~m
prevents sensitivity to vortex oscillations (Karman
eddies). Consequently, the additional tubes need no rubber dampers as
the ones presently mounted between the existing tubes.  The additional
triangle and two of the four vertical posts require together four
additional concrete foundation blocks of $1.5 \times 1.5$~m$^2$
surface. The other two vertical posts rest on the existing foundation
blocks.

The optical layout of the Option~II setup ($D\!=\!2.5$~m, $f\!=\!4.1$~m)
is shown in Fig.~\ref{rr-rh-fig6:optical layout II-IV} and is similar
to Option~I ($D\!=\!1.4$~m, $f\!=\!2.3$~m).  Both designs have an opening
ratio $f/1.64$. The 2.5~m version has the same important advantages:
no coma due to the use of two parabolic mirrors in cascade, low
polarization, user-selectable choice between angular resolution and
field size, possibility of transversal pupil shift and accompanying
choice of obstruction-free aperture, in this case up to 116\,cm.

The existing telescope mount has a central hole through its hour-angle
shaft.  In the Option~II design we propose to let the secondary beam
pass through it to a mirror which reflects the beam into a vertical
shaft to an optical lab on the ground.  In such a transfer system we
suggest to use a tandem AO system: the first to correct seeing-imposed
wavefront deformations and remaining tip-tilt fluctuations due to the
telescope structure, the second for the beam part along the hour-angle
axis, through the vertical shaft and in the optical lab.  An
alternative is to split the AO into a system for instrumentation
mounted on the telescope and an independent system for instrumentation
in the optical lab.

In the vertical shaft the light should travel parallel over the whole
length of the shaft, or large parts of it, in order to eliminate image
motion from wind-induced parallel translations between the platform
and the ground, which are of the order of 0.1~mm.  What remains are
small transversal shifts of the optical surfaces following wind gusts
and thus slow compared with the seeing motions. Such small pupil
shifts do not disturb the AO correction. In order to accommodate a
parallel beam over the whole shaft length without use of relay optics,
the minimum shaft diameter is $d = \sqrt{2\ell~D~\alpha} = 330$~mm
when a pupil image is located at the middle of the shaft for the
following parameter values: transfer length $\ell = 15000$~mm, primary
mirror diameter $D = 2500$~mm, field of view $\alpha =
5\pi/(180\times60)~{\rm rad} = 5$~arcmin, and pupil diameter $d/2 =
165$~mm.  For a pupil image at one end of the shaft, the diameters of
shaft and pupil become $\sqrt{2}$ times larger, 467 and 234~mm
respectively.  We propose to evacuate the vertical shaft using
entrance and exit windows to minimize the internal seeing.  The above
quantification shows that windows of only 50-cm diameter permit
transfer of a diffraction-limited image with $5\arcmin$ field of view
without requiring relay optics in the shaft.

There are other optical setups possible that avoid image motion from
parallel translation between platform and the ground level. An
example is one-to-one image transfer from the top to the bottom with
a relay lens halfway up the shaft which shifts over half the value
of the platform translation.  This can be realized with a passive
mechanical construction. The minimum shaft diameter then becomes two
times smaller, 165~mm instead of 330~mm.  Such a setup may fit
better in the overall optics layout.

\begin{figure}
  \centering \includegraphics[width=\textwidth]{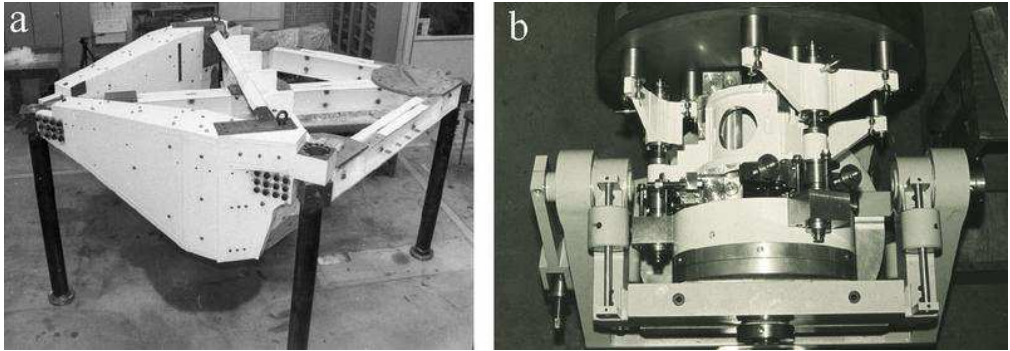}
  \caption[]{\label{rr-rh-fig7:platformassembly+support}
  DOT  inner platform and mirror support.
  \\{\em (a)\/} The existing inner platform of the DOT is an
  important complementary part to the equatorial mount and can be
  re-used together with the mount, also in the broader tower of
  Options IV \& V. This inner platform is shown here during the
  assembly at the TU workshop in Delft. \\{\em (b)\/} DOT mirror
  with its support in test mount. The 3-dimensional design of
  the whiffle-trees provides high stability against the varying wind
  load.
}
\end{figure}

The relay optics on top of the vertical shaft can and should be
connected to the inner platform of the DOT tower in a very stiff
way. The inner platform is an important complementary part to the
equatorial mount.  Fig.~\ref{rr-rh-fig7:platformassembly+support}a
shows this inner platform during its assembly at the university
workshop in Delft.  There are three connection plates for the
telescope mount: one on the corner point on the left side, the two
others are where the hoist eyebolts are placed on the diagonal from
the front to the back in this photograph. This inner platform
transforms in an extremely stiff way all forces and moments from the
telescope mount to direct forces without moments in its corner points,
to which the long downward tower tubes are connected directly.

Another very important mechanical part is the support of the primary
mirror. In the open concept, the support has to give stability against
the varying wind load on the primary mirror. For the 45-cm primary of
the present DOT a support system was developed that gives extreme
stability using only passive means.  A key aspect is the 3-dimensional
design of its individual support elements, including the whiffle trees
seen in Fig.~\ref{rr-rh-fig7:platformassembly+support}b.  This support
system is actually much stiffer than was required for the 45-cm
mirror.  A 2.5-m mirror support can consist of many more support units
of similar design. The mirror will be a classical thick mirror, but
hollow with a honeycomb structure with triangular cavities for high
stiffness.  Such honeycomb structures can be realized in common mirror
materials like Zerodur, Cervit or ULE, or a new material like Cesic
(SiC).  The stiff whiffle tree design avoids deformations by variable
gravity and wind loads to the extent that the wind pressure is
homogeneous over the mirror surface.  Only wind inhomogeneities
smaller than the mirror diameter, set by shear in the wind field,
cause moments that are not carried off by the whiffle trees. These
moments are at most a few tens of Nm and are carried by the thick
mirror structure. The thickness of the mirror will be larger than the
diameter of the AO sub-apertures projected to the primary
mirror. Consequently, the deformations by the wind-field shear are
small for the AO system due to the high bending stiffness of the thick
mirror.

Of course, many variants are possible for the optical layout.  The use
of focal-plane instrumentation directly on the pointed telescope
structure itself remains possible as well. In fact, the existing DOT
multi-wavelength imaging system could represent the initial instrument
package to start observing as in Fig.~\ref{rr-rh-fig3:optics1.4}a,
retaining its advantages of absence of image rotation and a relatively
simple optical scheme with few components.

\subsection{Option~III: upgrade to 2.5-m aperture on doubled tower}
\label{dot-sec:upgrade30mtower}

Option~III doubles the DOT tower in height in order to improve the
best-seeing occurrence frequency. At La Palma the frequency of
excellent seeing observations increases with the telescope height
above ground, roughly doubling at double height.  This was ascertained
in the ATST site selection surveys including La Palma under the
supervision of M.~Collados (IAC). Details are given in the final
report of the ATST Site Survey Working Group\footnote{ATST Project
  Documentation Report \#0021 Revision A, available at
  \url{http://atst.nso.edu/site/reports/} in file {\tt RPT-0021.pdf};
  in particular Appendix ''La Palma Height Comparison'' in files {\tt
  A13.11\_LaPalmaHtComp\_Final.pdf} and {\tt twoheight\_r0.pdf}.  The
  second file provides additional data to the first one.}.
The DOT is situated eleven meters lower than the SST. This is not a
disadvantage because the ATST test measurements show that what counts
is not the absolute height but the height above the local terrain; the
ground-layer turbulence follows its slope.  The DOT indeed experiences
closely similar seeing to the only slightly higher SST and its SVST
predecessor.

The tower height doubling is accomplished in Option~III by placing
Option~II (the existing tower with enlarged platform and additional
support carrying a 2.5-m telescope and a GREGOR-copy 9-m canopy) as
``upper'' tower on top of a new ``lower'' one.  The latter is again an
open steel framework with a geometry which maintains parallel platform
motion.  Fig.~\ref{rr-rh-fig5:upgrade2.5}c,d illustrates this doubling
to 30-m height with the telescope in observing and parked position,
respectively.

The principle of maintaining parallel platform motion in strong-wind
buffeting works as follows.  A horizontal wind load on the top
platform causes at the ``first floor'' level between the upper and lower
parts a horizontal force of the same strength and in the same
direction, plus a moment given by the force multiplied by the height
(15~m) of the upper part.  The direction of the moment is horizontal
and perpendicular to the direction of the wind force.  The four
original DOT-tower triangles and the fifth triangle added through
Option~II transfer this horizontal force and moment to the
intermediate first-floor level at which a complete network of steel
triangles ensures stiffness between the foot points of the five
upper-part triangles and four vertical posts.

The proposed lower-part framework between the base at ground level and
the first floor consists of five isosceles triangles with exactly the
same geometry as in the upper part.  In addition, each foot point of
the upper-part triangles and vertical posts is connected by a vertical
post to the ground level. The diameter of these fourteen posts is the
same as for the four in the top part, viz.\ 406.4~mm.  The five
lower-part triangles transfer the horizontal load downward from the
first floor to the base while keeping these two parallel to each
other.  The moment caused by the horizontal force on the top platform
is transported downward from the first-floor level by the vertical
posts attached to the foot points of the triangles of the upper part
of the tower.  Each moment-component consists of a tensile force and a
compressive one of equal value at the two foot points of an isosceles
triangle of the top part of the tower. The elongation and shortening
of the two vertical posts under the two foot points are equal in
value.  Consequently, the top point of the isosceles triangle in the
upper tower part moves horizontally, maintaining the parallel motion
of the top platform.

Elongation and shortening, respectively, of the two lower-part
vertical posts under the foot points of an upper-part triangle tilt
the latter at the first-floor level.  However, such tilts do not
influence the top platform because the upper-part triangles have
separate endpoints at the first floor.  All connections in this
floor are horizontal, perpendicular to the vertical posts in the
lower part. These connections cannot transport a significant
vertical force from one post to another because the bending
stiffness is much smaller than the tensile and compressive
stiffness.

The vertical component of a wind force on the top platform is
transported in a very stiff manner to the first floor by the
triangles and posts in the upper tower part. The fourteen posts in
the lower tower part transmit these forces directly and hence also
in an extremely stiff way to the foundation at ground level.

The horizontal wind forces on the first floor are brought downward
by the five isosceles triangles of the lower tower part while
keeping the first floor parallel to the ground floor, as explained
above.

All together, the top platform remains parallel to the ground under
wind loads on the tower. The existing concrete foundation can be
maintained while doubling the tower height to 30~m because its 2-m
deep concrete blocks are heavy enough to withstand the moments imposed
by any storm. In addition, their configuration with the lower-part
triangles anchored in separate blocks also favors holding the first
floor and top platform parallel to the ground.

The setup with an instrument lab at ground level fed by a vertical
vacuum tube remains as in Option~II.  In the case of parallel-beam
image transport, the doubling of the tube length necessitates
increase of the pupil and shaft diameters with $\sqrt{2}$, see
above.  The parallel translations increase from 0.1~mm to about
0.4~mm but this increase can be reduced through larger wall
thickness for the lower-part tubes.  One-to-one imaging from the top
to the bottom can be realized by two relay lenses, midway between
platform and the first level and midway between the latter and the
ground level, with an intermediate image with field lens located at
the first level. This setup offers the possibility of passive
mechanical relay lens translation to cancel the image motion at the
ground level. The minimum diameter of the shaft then remains 165~mm.

The existing electric elevator can be extended over the full height
and then connects the optical lab directly with the platform. The
needed space is available in the lower part while the upper part
retains the existing open elevator shaft.  Further options are closing
the cage, or possibly the whole elevator shaft, for easier platform
access in bad weather.

Note that a 30-m high tower of concrete cannot meet the stringent
demand on absence of platform tilt under large wind loads unless it is
much wider than this open-tower design employing tilt cancelation
through special geometry.  A wide concrete building would present a
severe obstacle to the ambient flow and jeopardize the seeing quality
at the telescope height.  The concrete SST tower is relatively slender
(narrower than its platform) and puts the SST imaging element (its 1-m
objective) high up in the wind.  In this sense even the evacuated SST
is an open telescope.

\begin{figure}
  \centering
  \includegraphics[width=\textwidth]{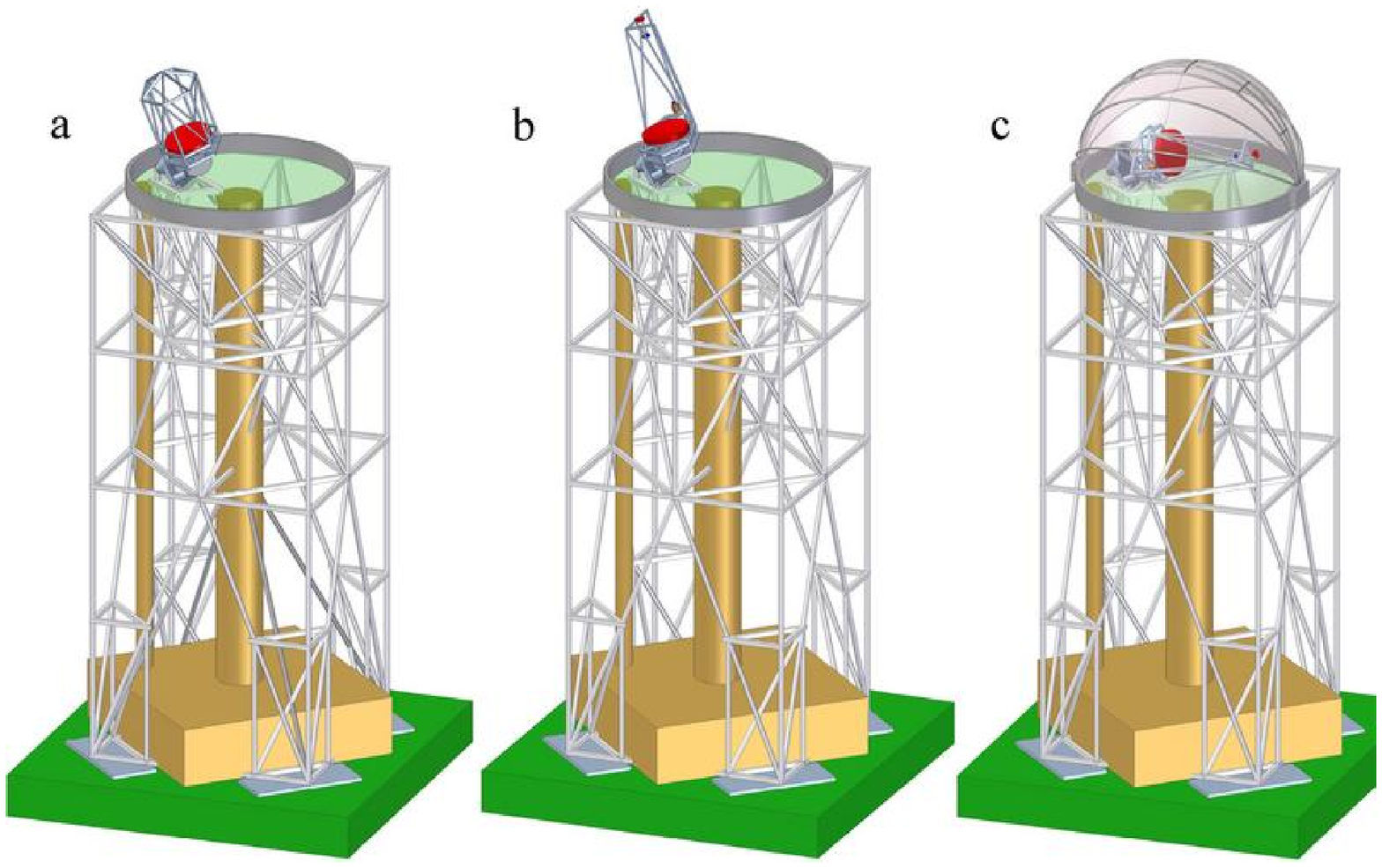}
  \caption[]{\label{rr-rh-fig8:upgrade2.5+newtower}
  Options~IV and V with 2.5-m primary mirror. \\{\em (a)\/} Option IV:
  new 30-m tower with the 2.5-m telescope of Options II
  and III.  The canopy has a diameter of 12~m.  The design for
  parallel motion of the platform is based on the principle of an
  outer and inner tower, see text. \\{\em (b)\/} Option V: 2.5-m
  telescope with off-axis primary mirror on the new tower. \\{\em
  (c)\/} Option V with closed canopy.
}
\end{figure}

\subsection{Option~IV: upgrade to 2.5-m aperture
with wide 30-m tower}  \label{dot-sec:upgradebroadert}

In Option~IV the existing equatorial DOT mount and inner platform are
to be placed on a new 30-m tower with a larger platform and a folding
canopy of 12~m diameter (Fig.~\ref{rr-rh-fig8:upgrade2.5+newtower}a)
in order to gain working space around the telescope.

The broader 30-m tower is again designed in a geometry which keeps
the top platform parallel to the ground under varying wind loads.
This version is based on the principle of having an outer and inner
double tower.  The outer tower consists of four 30-m vertical posts
holding the top platform parallel to the ground.  The inner tower
consists again of triangle framework providing high translational
stiffness as well as stiffness against rotations around a vertical
axis.  All connections between the inner and outer tower are
horizontal and hence perpendicular to the four outer posts so that
they do not disturb the parallel motion. The inner telescope
platform is connected to the four post tops with a separate
framework hanging free of the inner tower.  A more detailed
description of this tower design is given in
  \citet{rh-DOT-2006SPIE.6273E..50H}.
Its characteristics were checked through simulations with
finite-element analysis (program ANSYS,
  \cite{rh-DOT-2006SPIE.6268E..36L}).

The post-focus beam travels down through an evacuated shaft to an
optical lab on the ground, similar to the Option-III design.  In the
center of the tower an elevator shaft can be placed as indeed added
in Fig.~\ref{rr-rh-fig8:upgrade2.5+newtower}. The 12-m canopy would
be a further development of the 9-m GREGOR canopy. The tower can be
placed on a foundation of 4 blocks of concrete, each with a surface
of 3 $\times$ 5~m.

The tower shown in Fig.~\ref{rr-rh-fig8:upgrade2.5+newtower}a--c
implements the concept for a 30-m tower. The design principle of an
outer tower to hold the platform parallel and an inner tower for
stiffness remains suitable for yet higher towers: the shown
four-level geometry is suitable for heights up to 60\,m, while
larger height requires more stories.

\subsection{Option~V: upgrade to 2.5-m off-axis aperture on 30-m tower}
\label{dot-sec:upgradeoffax}

\begin{figure}
  \centering
  \includegraphics[width=\textwidth]{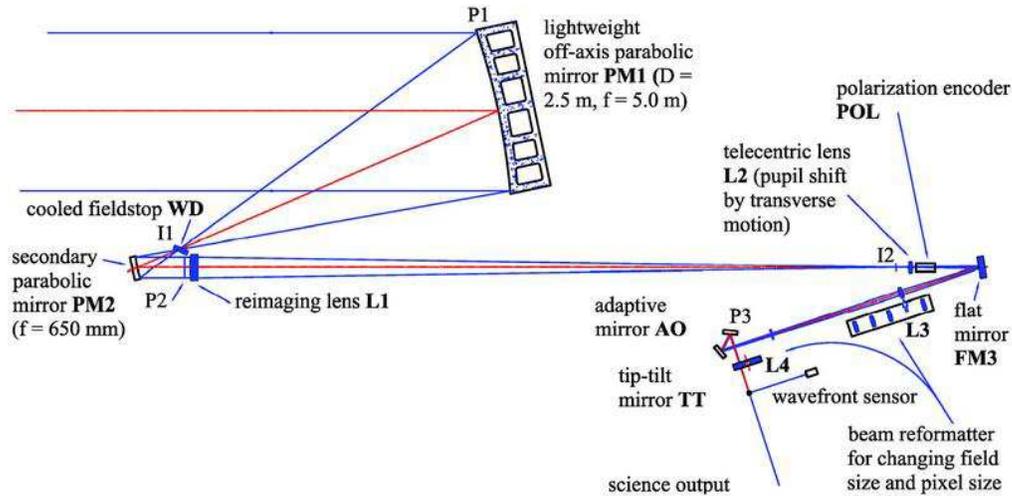}
  \caption[]{\label{rr-rh-fig9:optical layout V}
  Optical layout of
  Option~V with 2.5-m off-axis primary mirror.
}
\end{figure}

In this final and most expensive option an off-axis parabolic primary
mirror is purchased to obtain a fully clear aperture.
Figure~\ref{rr-rh-fig8:upgrade2.5+newtower}b,c shows such a telescope
placed on the new 30-m tower of Option~IV, in observing
position and in parked position with the folding canopy closed.
Figure~\ref{rr-rh-fig9:optical layout V} shows the optical layout.
The focal length of the primary mirror is 5~m.  The distance between
the optical axis and the nearest rim of the primary mirror is one
third of its diameter, hence 0.833~m.  A user-selectable choice
between angular resolution and field size and the possibility of
transversal pupil shift are again incorporated with a system similar
to that in the on-axis design.  For the off-axis design the
transversal pupil shift provides the option to use a smaller pupil
close to the optical axis to reduce partial polarization.

\begin{figure}
  \centering
  \includegraphics[width=\textwidth]{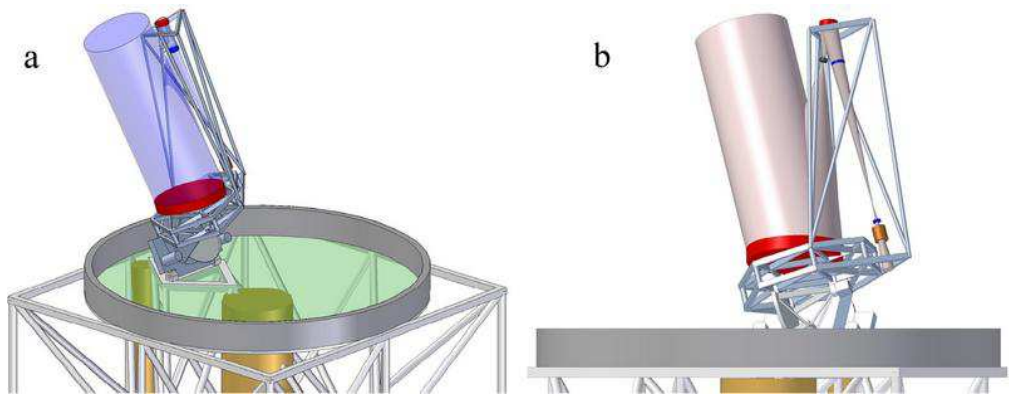}
  \caption[]{\label{rr-rh-fig10:offaxis}
  Off-axis design with stiff geometry of the framework supporting the
  secondary mirror in the telescope top.  \\{\em (a)\/} View from above.
  \\{\em (b)\/} View from behind, showing the framework geometry more
  clearly.  The optical beams are drawn as cylinders and cones. The
  beam from the primary mirror to the primary focus passes under the
  diagonal tube in the framework.
}
\end{figure}

The choice between on-axis and off-axis design has many aspects and
tends to invite many different opinions.  Mechanically, both are
possible.  Figure~\ref{rr-rh-fig10:offaxis}a,b shows in more detail
how also in an off-axis setup the mechanical structure of the
telescope top can be designed to guarantee stiff support of the
secondary mirror.  Optically, the off-axis design requires extremely
precise alignment between the primary and the secondary mirrors
transverse to the optical axis because otherwise the coma correction
becomes incomplete.

\section{Cost Estimates} \label{sec:costs}

Table~\ref{rr-rh-tab:costest} gives our cost estimates of the various
upgrade options. These do not include the purchase of the primary
mirror, nor the cost of post-focus instrumentation, nor the effort and
running costs of the DOT team.

The price of a primary mirror depends on market constraints such as
the order book and capacity of the few firms capable of fabrication
of large optics, for example the availability of a blank, the
occupancy of large polishing machines, etc.  We note that although
SiC is the preferential mirror material for large solar telescopes
particularly because of its excellent thermal conductivity (60 times
higher than Zerodur), the fully open designs proposed here can live
with Zerodur because their mirror flushing becomes sufficiently
effective already at medium wind speeds.  The larger weight of
Zerodur poses no problem in these designs because the DOT mount,
re-used in all of them, is sufficiently over-dimensioned (see
above).

The cost of post-focus instrumentation obviously ranges from low
(re-usage of the existing DOT multi-camera multi-wavelength imaging
system) to large (the image lab at ground level harboring many
instruments).  We envisage emphasis on high-cadence line-profile
sampling with LC polarimetry for large fields of view and
two-dimensional multi-line spectrometry including spectropolarimetry
with phase-diverse MOMFBD restoration using a fiber field-of-view
reformatter from two-dimensional to slit geometry
  (cf.\ \cite{rh-Rutten1999c}). 

The DOT team is indispensable in the proposed upgrades not only
because the latter use bits and pieces of the present DOT but yet
more because of the team's expertise in tower and telescope
open-design principles, computational verification of these,
hardware solutions, and the La Palma environment.  In addition, the
teams would supply the intensive contacts with the university
workshops (at Utrecht and Delft) whose contributions are budgeted
separately in the next-to-last column in
Table~\ref{rr-rh-tab:costest}.  A rough estimate of the DOT-team
manpower, management, and operational costs to undertake the
proposed upgrades ranges from 250 kEuro/year (the present DOT
operations budget) to 500 kEuro/year.

\begin{figure}
  \centering
  \includegraphics[width=\textwidth]{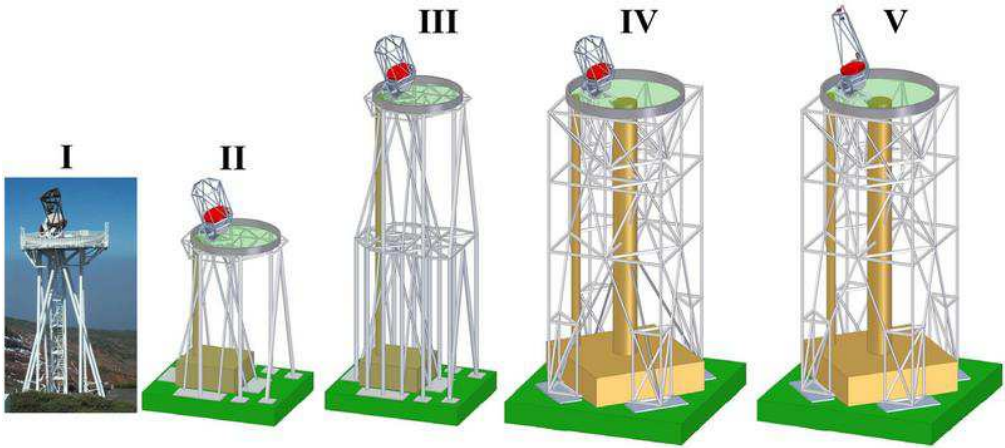}
  \caption[]{\label{rr-rh-fig11:overview}
  Overview of the five options.  I: upgrade of the present DOT to
  1.4-m aperture.  II--IV: 2.5~m aperture with increasing tower height
  and platform width.  V: 2.5~m off-axis version.
}
\end{figure}

\begin{table}
  \begin{center}
  \caption[]{\label{rr-rh-tab:costest}
  Cost estimates in kEuro, excluding the primary
  mirror, post-focus instrumentation, and DOT-team effort.}
  \vspace*{0.5ex}
  {\begin{tabular}{ccccrcr}
  \tableline
  \noalign{\smallskip}
  Option$^{*}$ & Optics & Telescope & Tower+ & Canopy & University &
  Total  \\
    &  & structure & platform & & workshops &  \\
  \noalign{\smallskip}
  \tableline
  \noalign{\smallskip}
  I   & 200 & 300 &      &      & 300 &  800 \\
  II  & 300 & 400 &  300 &  600 & 300 & 1900 \\
  III & 400 & 400 &  900 &  600 & 600 & 2900 \\
  IV  & 400 & 400 & 3200 & 1100 & 600 & 5700 \\
  V   & 600 & 400 & 3200 & 1100 & 600 & 5900 \\
  \noalign{\smallskip}
  \tableline
  \end{tabular}
  }
  \end{center}

 \begin{small}
 \begin{tabbing}
  $^{*}$ \= I \hspace{3mm} \= 1.4-m aperture with existing 7-m
  canopy and existing 15-m tower \\
    \> II \> 2.5-m aperture with 9-m GREGOR-like canopy and
             widened 15-m tower \\
    \> III \> 2.5-m aperture as Option II with a doubled 30-m tower\\
    \> IV \> 2.5-m aperture with 12-m canopy and new, wide 30-m tower\\
    \> V \> the same as Option~IV with a 2.5-m off-axis primary mirror\\
 \end{tabbing}
 \end{small}
\end{table}

The costs listed in Table~\ref{rr-rh-tab:costest} under ``Optics''
cover the prime-focus and subsequent packages.  ``Telescope
structure'' includes a new mirror support.  The University workshops
are needed for design work on the telescope structure, tower and
platform, in addition to the industrial manufacturing costs listed in
the respective columns.  Special small parts and test equipment must
also be produced in these workshops. The increase in workshop costs
from Option~II to Option~III reflects additional design work for the
higher towers, in particular the far-from-trivial joints in the corner
points, the beam transfer tube, and the elevator.  The joints need
very careful designing in order to maintain the high stiffness of the
overall geometry, and will need complex special parts that are best
manufactured in-house.

Figure~\ref{rr-rh-fig11:overview} shows a pictural overview of the five
options.  In summary: without inclusion of the DOT team, the primary
mirror, and post-focus instrumentation, Option~I takes roughly
1~M{Euro}, Option~II 2~M{Euro}, Option~III 3~M{Euro}, and options IV
and V 6~M{Euro}.  The latter two options are costlier due to their new
tower and foundation.

These various options may be realized in overlapping phases with
continued observing.  For instance, the Option-II platform and canopy
enlargements can be realized keeping the DOT in operation, with fast
subsequent upgrading of the telescope.  The Option-III tower doubling
may follow on the Option~II aperture upgrade.  In Options~III--V the
higher tower may be realized first and tested with the existing
telescope before the latter is replaced.  Other variants and option
combinations are also possible.

\section{Conclusion}

The DOT telescope and tower combine rigorously open design with
extraordinary stiff construction.  This combination was achieved
through the application of special geometries and the careful
attention to the design of joints and drives that is necessary to
indeed realize the geometrical stiffness.  The excellent DOT
high-resolution observations have amply proven both the open concept
and the DOT's pointing stability.

We would like to enlarge the DOT and profit fully from its concept.
Only relatively moderate investments are needed through re-using
existing DOT parts, in particular the equatorial mount and drives
which are indeed oversized for the present 45-cm primary mirror.  In
this paper we have outlined strawman designs for upgrades to 1.4-m and
2.5-m aperture, optionally adding a groundbased post-focus lab and
30-m tower height to double the frequency of super-seeing.  Even with
a 2.5-m primary mirror diffraction-limited pointing stability can be
reached without compromising the rigorously open construction of both
the telescope and the tower.

We welcome partners to share in DOT operation and the realization of
one of these upgrades.\\

\acknowledgements We thank the CSPM organisers for a very good
  meeting.  The DOT is presently operated by Utrecht University at the
  Spanish Observatorio del Roque de los Muchachos of the Instituto de
  Astrof{\'{\i}}sica de Canarias with funding from Utrecht University
  and the Netherlands Graduate School for Astronomy NOVA.  
  A.\,P.\,L. J\"agers is supported by the Netherlands Technology Foundation
  STW.  R.\,J. Rutten acknowledges travel support from the Leids
  Kerkhoven-Bosscha Fonds and the European Solar Magnetism Network.


\begin{thebibliography}{}

\bibitem[\protect\astroncite{Bettonvil et~al.}{2004}]{rh-DOT-dot++2004}
Bettonvil F. C.~M., Hammerschlag R.~H., S{\"u}tterlin P., Rutten R.~J.,
  J{\"a}gers A.~P., Snik F., 2004,
\newblock in J. Oschmann (ed.), Astronomical Telescopes and Instrumentation,
  Procs.\ SPIE 5489,
  362

\bibitem[\protect\astroncite{{Bettonvil}
  et~al.}{2006}]{rh-DOT-2006SPIE.6269E..12B}
{Bettonvil} F.~C.~M., {Hammerschlag} R.~H., {S{\"u}tterlin} P., {Rutten} R.~J.,
  {J{\"a}gers} A.~P.~L., {Sliepen} G., 2006,
\newblock in I.~S. McLean, M. Iye (eds.), Ground-based and Airborne
  Instrumentation for Astronomy, Procs.\ SPIE 6269,
   paper 62690E

\bibitem[\protect\astroncite{Hammerschlag}{1981}]{rh-DOT-Hammerschlag1981a}
Hammerschlag R.~H., 1981,
\newblock in R.~B. Dunn (ed.), Solar Instrumentation: What's Next?, Proc.\
  Sacramento Peak Nat'l Obs.\ Conf.,
  Sunspot, New Mexico,
  547

\bibitem[\protect\astroncite{Hammerschlag}{1983}]{rh-DOT-Hammerschlag1983}
Hammerschlag R.~H., 1983,
  Procs.\ SPIE~  444, 138

\bibitem[\protect\astroncite{{Hammerschlag}
  et~al.}{2006a}]{rh-DOT-2006SPIE.6273E..50H}
{Hammerschlag} R.~H., {Bettonvil} F.~C.~M., {J{\"a}gers} A.~P.~L., 2006a,
\newblock in E. Atad-Ettedgui, J. Antebi, D. Lemke (eds.), Optomechanical
  Technologies for Astronomy, Procs.\ SPIE 6273,
   paper 62731O

\bibitem[\protect\astroncite{{Hammerschlag}
  et~al.}{2006b}]{rh-DOT-Hammerschlag++Beijing2006}
{Hammerschlag} R.~H., {Bettonvil} F.~C.~M., {J{\"a}gers} A.~P.~L., {Sliepen}
  G., {Snik} F., 2006b,
\newblock in Spatial Structures, Procs.\ IASS,
   paper 108

\bibitem[\protect\astroncite{{Keller} \& {von der
  L{\"{u}}he}}{1992}]{rh-1992A&A...261..321K}
{Keller} C.~U., {von der L{\"{u}}he} O., 1992,
  \aap~  261, 321

\bibitem[\protect\astroncite{{Lanford} et~al.}{2006}]{rh-DOT-2006SPIE.6268E..36L}
{Lanford} E., {Swain} M., {Meyers} C., {Muramatsu} T., {Nielson} G., {Olson}
  V., {Ronsse} S., {Vinding Nyden} E., {Hammerschlag} R., {Little} P., 2006,
\newblock in {Monnier, John D.}, {Sch{\"o}ller, Markus}, {Danchi, William C.}
  (eds.), Advances in Stellar Interferometry, Procs.\ SPIE 6268,
   paper 626814

\bibitem[\protect\astroncite{Rutten}{1999}]{rh-Rutten1999c}
Rutten R.~J., 1999,
\newblock in T.~R. Rimmele, K.~S. Balasubramaniam, R.~R. Radick (eds.), High
  Resolution Solar Physics: Theory, Observations, and Techniques, Procs.\ 19th
  NSO/Sacramento Peak Summer Workshop,
  \aspcs,
  Vol.~183,
  296

\bibitem[\protect\astroncite{Rutten et~al.}{2004a}]{rh-DOT-StPetersburg2004}
Rutten R.~J., Bettonvil F. C.~M., Hammerschlag R.~H., J{\"a}gers A. P.~L.,
  Leenaarts J., Snik F., S{\"u}tterlin P., Tziotziou K., de~Wijn A.~G., 2004a,
\newblock in A.~V. Stepanov, E.~E. Benevolenskaya, A.~G. Kosovichev (eds.),
  Multi-Wavelength Investigations of Solar Activity, Procs.\ IAU Symposium 223,
  Cambridge University Press,
  597

\bibitem[\protect\astroncite{Rutten et~al.}{2004b}]{rh-DOT-tomo1}
Rutten R.~J., Hammerschlag R.~H., Bettonvil F. C.~M., S{\"u}tterlin P., de~Wijn
  A.~G., 2004b,
  \aap~  413, 1183

\bibitem[\protect\astroncite{Snik et~al.}{2007}]{rh-DOT-SPW4}
Snik F., Bettonvil F. C.~M., J{\"a}gers A. P.~L., Hammerschlag R.~H., Rutten
  R.~J., Keller C.~U., 2007,
\newblock in B.~W. Lites, R. Casini (eds.), Procs.\ 4th Solar Polarization
  Workshop, \aspcs,
   in press

\bibitem[\protect\astroncite{{van Noort} et~al.}{2005}]{rh-2005SoPh..228..191V}
{van Noort} M., {Rouppe van der Voort} L., {L{\"o}fdahl} M.~G., 2005,
  \solphys 228, 191

\bibitem[\protect\astroncite{{van Noort} \& {Rouppe van der
  Voort}}{2006}]{rh-2006ApJ...648L..67V}
{van Noort} M.~J., {Rouppe van der Voort} L.~H.~M., 2006,
  \apjl~  648, L67

\end{thebibliography}

\end{document}